%% file: conf_sum2008.tex
\documentclass[11pt]{article}

\usepackage{graphicx}
\usepackage{subfigure}

\newcommand{\BABARPubYear}    {08}

\newcommand{\BABARConfNumber} {006}
\newcommand{\SLACPubNumber} {13350}

\input{babarsym}
\input{extrababarsym}

\long\def\inst#1{\par\nobreak\kern 4pt\nobreak
    {\it #1}\par\vskip 10pt plus 3pt minus 3pt}

\begin{document}
{\pagestyle{empty}

\begin{flushright}
\babar-CONF-\BABARPubYear/\BABARConfNumber \\
SLAC-PUB-\SLACPubNumber \\
arXiv:0808.1121 [hep-ex] \\
July 2008 \\
\end{flushright}

\par\vskip 5cm

\begin{center}
\Large \bf Measurement of $\mathcal{B}(\tau^{-}\rightarrow \bar{K^{0}}\pi^{-}\nu_{\tau})$ using the $\babar$ detector 
\end{center}
\bigskip

\begin{center}
\large The \babar\ Collaboration\\
\mbox{ }\\
\today
\end{center}
\bigskip \bigskip

\begin{center}
\large \bf Abstract
\end{center}
A preliminary measurement of the branching fraction $\mathcal{B}(\tau^{-}\rightarrow K^{0}_{S}\pi^{-}\nu_{\tau})$ is
made using $384.6 \invfb$ of
\epem\ collision data provided by the \pep2\ collider, operating primarily at $\sqrt{s}=10.58 \gev$,
and recorded using the \babar\ detector.
From this we measure: $\mathcal{B}(\tau^{-}\rightarrow \bar{K^{0}}\pi^{-}\nu_{\tau}) = (0.840 \pm 0.004 \, \stat \pm 0.023 \, \syst) \%.$
This result is the most precise measurement to date and is consistent with the world average.

\vfill
\begin{center}

Submitted to the 33$^{\rm rd}$ International Conference on High-Energy Physics, ICHEP 08,\\
30 July---5 August 2008, Philadelphia, Pennsylvania.

\end{center}

\vspace{1.0cm}
\begin{center}
{\em Stanford Linear Accelerator Center, Stanford University, 
Stanford, CA 94309} \\ \vspace{0.1cm}\hrule\vspace{0.1cm}
Work supported in part by Department of Energy contract DE-AC02-76SF00515.
\end{center}

\newpage
} 

%
\input{authors_ICHEP2008.tex}

\section{INTRODUCTION}
\label{sec_introduction}
\input{sections/intro}

\section{THE \babar\ DETECTOR AND DATASET}
\label{sec:babar}
The \babar\ detector is described in detail in \cite{Aubert:2002XX}.
Charged particles are detected and their momenta measured with a 5-layer double
sided silicon vertex tracker (SVT) and a 40-layer drift chamber (DCH) inside a $1.5 \rm{\, T}$
superconducting solenoidal magnet. A ring-imaging Cherenkov detector (DIRC) is used
for the identification of charged particles. Energies of neutral particles are
measured by an electromagnetic calorimeter (EMC) composed of $6,580$ CsI(Tl) crystals,
and the instrumented magnetic flux return (IFR) is used to identify muons.

\input{sections/data}
\input{sections/montecarlo}

\section{SELECTING $\tau^{-}\rightarrow K^{0}_{S}\pi^{-}\nu_{\tau}$ EVENTS}\label{sec_kpi0eventselection}

\input{sections/selectionCuts}

\input{sections/sigEfficiency}

\section{EFFICIENCY CORRECTIONS}\label{sec:EffCorrections}
\input{sections/EfficiencyCorrections}

\section{SYSTEMATIC UNCERTAINTIES}\label{sec_systematics}
\input{sections/systematics}

\section{BRANCHING FRACTION MEASUREMENT: $\mathcal{B}(\tau^{-}\rightarrow \bar{K^{0}}\pi^{-}\nu_{\tau})$}\label{sec_brkpi0}
\input{sections/branchingRatio}

\section{SUMMARY}\label{sec_summary}
\input{sections/summary}

\section{ACKNOWLEDGEMENTS}\label{sec_acknowledgements}
\input{acknowledgements}

\end{document}

%% file: extrababarsym.tex
\def\Nsiggen                 {\ensuremath{N_{\scriptscriptstyle{\rm sig}}^{\scriptscriptstyle{\rm gen}}}\xspace}
\def\Nsigsel                 {\ensuremath{N_{\scriptscriptstyle{\rm sig}}^{\scriptscriptstyle{\rm sel}}}\xspace}

\def\Ntautau                 {\ensuremath{N_{\scriptscriptstyle{\tau\tau}}}\xspace}

\def\SigEff                  {\ensuremath{\varepsilon_{\scriptscriptstyle{\rm sig}}}\xspace}

\def\EffCorr                 {\ensuremath{\varepsilon_{\scriptscriptstyle {\rm corr}}}\xspace}

\def\taumtoKpim              {\ensuremath{\taum \to (\kaon \pi)^- \nu_\tau}\xspace}

\def\uds                     {\ensuremath{\u \d \s}\xspace}

\def\strangemass             {\ensuremath{m_{\scriptscriptstyle \s}}\xspace}

\def\Kstar                   {\ensuremath{K^{*}}\xspace}
\def\Kstarp                  {\ensuremath{K^{*+}}\xspace}

\def\Kstareightninetwom      {\ensuremath{K^{*}(892)^{-}}\xspace}

\def\Kone                    {\ensuremath{K_{1}}\xspace}

\def\Kpi                     {\ensuremath{\kaon \pi}\xspace}
\def\Kpipi                   {\ensuremath{\kaon \pi \pi}\xspace}

\def\Krho                    {\ensuremath{\kaon \rho}\xspace}
\def\Kstarpi                 {\ensuremath{\Kstar \pi}\xspace}

\def\cabibbo                 {\ensuremath{\theta_{\scriptscriptstyle C}}}

\def\suthreef                {\ensuremath{SU(3)_{f}}\xspace}

\def\dz                      {\ensuremath{d_{\scriptscriptstyle 0}\xspace}}

\def\aleph                   {ALEPH}
\def\cleo                    {CLEO}
\def\opal                    {OPAL}

\def\belle                   {BELLE}

\def\kk                      {{\tt KK2f}}
\def\tauola                  {{\tt TAUOLA}}

\def\ie                      {\mbox{i.e.}}

\def\ckm                     {CKM}

\def\etag                    {$e$-tag}
\def\mutag                   {$\mu$-tag}
\def\taubkgd                 {\mtau\ background}
\def\taubkgds                {\mtau\ backgrounds}

\def\av                      {axial-vector}

\def\mc                      {Monte Carlo}

\def\crosssection            {cross-section}


%% file: authors_ICHEP2008.tex
\begin{center}
\small

The \babar\ Collaboration,
\bigskip

%
B.~Aubert,
M.~Bona,
Y.~Karyotakis,
J.~P.~Lees,
V.~Poireau,
E.~Prencipe,
X.~Prudent,
V.~Tisserand
\inst{Laboratoire de Physique des Particules, IN2P3/CNRS et Universit\'e de Savoie, F-74941 Annecy-Le-Vieux, France }
J.~Garra~Tico,
E.~Grauges
\inst{Universitat de Barcelona, Facultat de Fisica, Departament ECM, E-08028 Barcelona, Spain }
% ...

\end{center}\newpage

%% file: sections/intro.tex
Hadronic \mtau\ decays provide a clean laboratory for studying the hadronic weak current.
Hadronic products from \mtau\ decays give access to the 
light quark vector \ensuremath{(V)} and \av\ \ensuremath{(A)} spectral functions, which give insight into 
the dynamics of QCD at intermediate scales as well as providing tests of the Standard Model itself 
\cite{Glashow:1961tr}.   

For hadronic $\tau$-decays, \suthreef\ 
symmetry breaking can be used to determine the Cabibbo-Kobayashi-Maskawa (\ckm) matrix 
element magnitude \Vus \cite{Cabibbo:1963yz}, the strong coupling constant, \as, and the strange 
quark mass, \strangemass~\cite{Jamin}.  These are all tests of the Standard Model as deviations from 
values measured in other processes would indicate new physics.

Hadrons from \mtau\ decays are produced via \W\ emission.  Relative to non-strange (\ensuremath{\u \d}) currents, 
strange (\ensuremath{\u \s}) currents of \mtau\ decays are suppressed by \ensuremath{\left( \Vus / \Vud \right)^{\scriptscriptstyle 2} \simeq \tan^{\scriptscriptstyle 2} \cabibbo}, where \Vud\ and \Vus\ are the absolute values of the \ckm\ matrix elements.  Resonant decay dominates these currents: the strange vector current is dominated by a \Kstar\ resonance 
which decays to \Kpi and the strange \av\ current by the \Kone\ which decays mostly via \Krho\ and \Kstarpi\ to \Kpipi.  

The high luminosity provided by the \pep2\ collider, coupled with a large $\tautau$ production \crosssection\ near 
the operating energy of $\sqrt{s}=10.58 \gev$, provides a large data sample with which to study the strange hadronic 
decay \taumtoKpim\ using the \babar\ detector.  Measurements of \mtau\ decay branching fractions to strange 
hadronic final states and studies of the strange spectral functions have been conducted by \aleph\ \cite{Chen:2001qf}, 
\cleo\ \cite{Briere:2003fr} and \opal\ \cite{Abbiendi:2004}, but have been limited by statistics.
About a hundred times larger samples of $\tau$-events have been provided by the $B$-factories \babar\ and \belle.    
The measurement of $\mathcal{B}(\tau^{-}\rightarrow K^{0}_{S}\pi^{-}\nu_{\tau})$ was recently carried out by \belle\ \cite{BellePaper} with a significantly reduced uncertainty compared to that of previous measurements.

%% file: sections/data.tex
The analysis described in this paper is based on data taken using the \babar\ detector
at the \pep2\ collider \cite{SLAC-372b} located at SLAC in the data-taking periods
between October 1999 and August 2006. During this period a total of $384.6 \invfb$ of data was recorded with a cross-section
for \tautau\ pair production of $(0.919 \pm 0.003) \nb$ \cite{Banerjee:2008}.  This data sample contains
over 700 million \mtau\ decays. 

%% file: sections/montecarlo.tex
\mc\ (MC) studies of simulated signal and background events were
carried out using various MC samples. The $\tau$ MC
events studied were generated with \kk~\cite{Ward:2002qq} and decayed with 
\tauola~\cite{tauola} using \mtau\ branching fractions based on PDG 2006~\cite{Yao:2006fs}.
In the MC, the $\taum$ decays to $K^{0}_{S}\pi^{-}$ via the \Kstareightninetwom\ resonance with a branching fraction
of 0.90\%. Non-\mtau\ hadronic and dilepton MC samples are used for studying the non-\mtau\ backgrounds.

%% file: sections/selectionCuts.tex
In this analysis events containing a pion and a $K_{S}^{0}$ in the final state are studied, where the $K_{S}^{0}$ is 
reconstructed in the $\pi^{+}\pi^{-}$ mode.

The event is divided into two hemispheres in the center-of-momentum system (CMS)
using the plane perpendicular to the thrust axis, which is the direction which maximizes the sum of the longitudinal momenta of 
the neutrals and tracks in the event \cite{thrust}. 
One hemisphere of the event is required to contain only one charged track; this is defined as the tag hemisphere.  
The other hemisphere is required to contain three charged tracks; this is defined as the signal hemisphere.  The tag 
track and at least one of the signal hemisphere tracks are required to point towards the interaction point. 

Approximately 35 \% of $\tau$ decay to fully leptonic final states.  
Requiring the track in the tag hemisphere to be identified as an electron or 
muon while requiring the signal hemisphere to contain only hadrons strongly reduces 
backgrounds from $e^{+}e^{-}\rightarrow q\bar{q}$ events.  Electrons are identified using the ratio 
of calorimeter energy to the track momentum ($E/p$), the
ionisation loss in the tracking system (\dedx) and the shape of the shower in the
calorimeter.  Muons are identified by hits in the IFR and small energy deposits in the calorimeter.

$K^{0}_{S}$ candidates are constructed from any two oppositely charged tracks with an invariant mass within 25
MeV/$\textrm{c}^{2}$ of the $K_{S}^{0}$ mass as given by the Particle Data Group (PDG) 
497.672~MeV/$\textrm{c}^{2}$~\cite{Yao:2006fs}.  Only events with exactly one $K^{0}_{S}$ candidate are retained.  
The track from the signal side not originating from the $K^{0}_{S}$ candidate is required to be identified as a 
pion and originate from the 
interaction point.  Pions are identified by
the ionisation loss in the tracking system (\dedx), the shape of the shower in the
calorimeter and their discrimination from kaons performed in the DIRC.  
  All tracks on the signal side are required to lie within the
geometrical acceptance region of the EMC and DIRC to ensure good particle identification.

Additional cuts are imposed to further reduce the backgrounds. The net charge of the event is required 
to be zero and a cut requiring the thrust of the event to be greater than 0.85 is imposed 
to reduce the non-\mtau\ background.

Backgrounds from Bhabha events are suppressed by requiring the momentum of the lepton-side track to be less than 
4.9~GeV/$\textrm{c}$.
Backgrounds from radiative Bhabha and $\mu$-pair events with a converted photon are suppressed by requiring the modulus of the cosine 
of the decay angle to be less than 0.97.  The decay angle is defined as the angle between the momentum of the 
$\pi^{+}$ originating from the $K^{0}_{S}$ in the $K^{0}_{S}$'s rest frame and the $K^{0}_{S}$ momentum in the 
laboratory frame.  When this quantity is calculated on electrons misidentified as pions the effect of assigning an 
incorrect mass will push the value towards $\pm$1.  From studies of missing transverse event energy, backgrounds
from two-photon events are determined to be negligible.

\begin{figure}[htbp]
\begin{center}
\includegraphics[width=27pc,height=22.5pc]{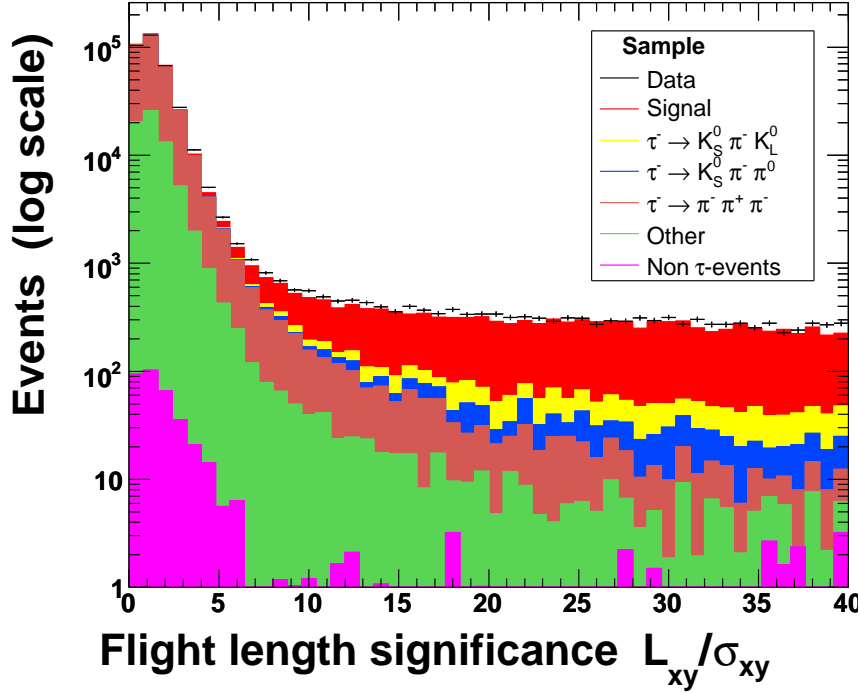}
\caption{\footnotesize{The $K_{S}^{0}$ flight length significance in the $xy$-plane for the the combined (\etag+\mutag) 
sample.  The MC signal branching fraction is set to the measured value from this analysis.}}
\label{fig:KsFliSig}
\end{center}
\end{figure}

Along with signal events, $\tau^{-}\rightarrow\pi^{-}\pi^{+}\pi^{-}\nu_{\tau}$ events also produce three pions in the final state, but all three 
come directly from the primary interaction point.  To remove $\tau^{-}\rightarrow\pi^{-}\pi^{+}\pi^{-}\nu_{\tau}$ 
events the $K^{0}_{S}$ 
flight length significance in the plane perpendicular to the collision axis is 
required to be greater than 5.0.  The flight length significance is defined as the measured flight length divided by its 
estimated uncertainty.  This cut removes approximately 90\% of the remaining 
$\tau^{-}\rightarrow\pi^{-}\pi^{+}\pi^{-}\nu_{\tau}$ events in the sample.  Figure\,\ref{fig:KsFliSig} shows 
the $K_{S}^{0}$ flight length significance distribution for events in this analysis with all other cuts applied.

To increase the likelihood that the pions from the $K^{0}_{S}$ 
candidate really come from a $K^{0}_{S}$ they are
required to have a distance of closest approach to each other of less than 2 mm.  
The $K_{S}^{0}$ trajectory is reconstructed by performing a vertex fit on the
$K_{S}^{0}$ daughter pions constraining them to originate from a single point and
then summing their momenta.  To increase the probability that the
$K^{0}_{S}$ originated from the interaction point this trajectory is
required to have a distance of closest approach from the collision
axis of less than 1 mm.

Once these cuts have been applied, the largest two background channels are from 
$\tau^{-}\rightarrow K^{0}_{L}K^{0}_{S}\pi^{-}\nu_{\tau}$ and
$\tau^{-}\rightarrow K^{0}_{S}\pi^{-}\pi^{0}\nu_{\tau}$ events where the additional neutral
particle to the $K^{0}_{S}$ is undetected.  In order to reject such events maximum neutral energy cuts are applied.
The total energy in the calorimeter not associated to any charged track must be less than 0.5 GeV.  
In the signal hemisphere this quantity is required to be below 0.25 GeV.

%% file: sections/sigEfficiency.tex
After selection about 80\% of the retained MC events are $\tau\rightarrow K^{0}_{S}\pi^{-}\nu_{\tau}$.  About 
98.5\% of the background events come from $\tau$-decays and about 1.5\% from non-$\tau$-events.   
The overall signal efficiency is estimated from the MC simulation as:
\begin{equation}
\SigEff = \frac{\Nsigsel}{\Nsiggen},
\label{eq:SigEff}
\end{equation}
where $\Nsiggen$ is the number of generated signal events and \Nsigsel\ is the number of
selected signal events.
A total of $5.872 \times 10^{\scriptscriptstyle 6}$ $\tau\rightarrow \bar{K^{0}}\pi^{-}\nu_{\tau}$ signal events 
were generated.

The total selection efficiency, which includes the efficiency corrections that are described in 
Section~\ref{sec:EffCorrections}, is $(0.620 \pm 0.003)\%$, $(0.482 \pm 0.003)\%$ and $(1.101 \pm 0.004)\%$ for the \etag, 
\mutag\ and combined samples respectively.

%% file: sections/EfficiencyCorrections.tex
Since imperfect detector simulation may mean that the reconstruction/selection efficiencies differ between data and \mc, 
some corrections are applied to \mc\ signal and background events.

A detailed study of the $K_{S}^{0}$ efficiency has been carried out. 
$B^{0}\rightarrow\pi^{+}D^{-}(D^{-}\rightarrow K^{0}_{S}\pi^{-})$ decays were used to 
study the $K^{0}_{S}$ efficiency for data and \mc.
As a result of this study, a per $K^{0}_{S}$ average data/MC efficiency 
correction of 0.983 should be applied
to the MC sample.
Since only one $K^{0}_{S}$ is selected in this analysis the MC sample is weighted with $0.983$ as a $K^{0}_{S}$ efficiency correction.

The efficiency of particle identification (PID) cuts are known to differ between data 
and \mc\ and so corrections to the \mc\ are applied.
A set of efficiency tables, binned in $p$, $\theta$ and $\phi$, are used in order to obtain the 
necessary weights to use to correct the \mc\ events. 
The efficiency correction (data/MC relative efficiency) and corresponding uncertainty is calculated for each 
track.

In this analysis charged PID selection is performed on the pion not originating from the $K^{0}_{S}$ and the lepton. 
The average values obtained for the corrections due to the electron, muon and pion are $1.015$, $0.948$ 
and $0.979$ respectively.

The total data/MC efficiency correction, \EffCorr, is made by combining the efficiency corrections described above.
\EffCorr\ is used to weight the MC sample and the average values obtained are $0.977$, $0.912$ and $0.947$
for the \etag, \mutag\ and combined samples.
The systematic error coming from this procedure is described in Section~\ref{sec_systematics}.

%% file: sections/systematics.tex
A number of systematic uncertainties have been considered.  These are described below.

An error of $1.10\%$ is included due to the $K^{0}_{S}$ efficiency correction procedure described in Section~\ref{sec:EffCorrections}.  
When propagated through to the branching fraction measurement this gives a $1.40\%$ systematic uncertainty on the 
branching fraction.

The tracking efficiency is susceptible to bias caused by physics data and MC simulation discrepancies.
An error of $0.23\%$ per track not originating from the $K^{0}_{S}$ is assigned to account for this.
Tracking uncertainties from the $K^{0}_{S}$ daughters are included in the $K^{0}_{S}$ selection systematic.
Each event contains two reconstructed charged tracks with correlated uncertainties, leading to a total tracking efficiency 
uncertainty of $0.46\%$.  This gives a $0.58\%$ systematic uncertainty on the branching fraction.

The total particle identification uncertainty that arises from the efficiency correction procedure described in 
Section~\ref{sec:EffCorrections} for the \etag, \mutag\ and combined samples is estimated to be $1.05\%$,
$1.33\%$ and $1.18\%$, respectively.  When propagated through to the branching fraction measurement this gives 
a systematic uncertainty of $1.45\%$ for the \etag, $1.68\%$ for the \mutag\ and $1.50\%$ for the combined sample.

The relative uncertainty associated with the \tautau\ pair production cross-section and the \babar\ luminosity 
determination is $0.65\%$, leading to a $0.83\%$ systematic uncertainty on the branching fraction.  A relative error is 
applied 
for uncertainty in the modelling of selection variables on which cuts are applied.  The modelling efficiency uncertainty 
is estimated as $0.29\%$, leading to a $0.37\%$ systematic uncertainty on the branching fraction.

The systematic uncertainty on the signal efficiency and branching fraction measurement due to signal \mc\ statistics 
is $0.51\%$ for the \etag, $0.66\%$ for the \mutag\ and $0.37\%$ 
for the combined samples. 
The error on the number of background events due to limited \mc\ statistics is $1.02\%$ for \etag, $1.12\%$ for \mutag\ 
and $0.76\%$ for the combined sample.  This gives a systematic uncertainty on the branching fraction measurement of
$0.28\%$ for \etag, $0.30\%$ for \mutag\ and $0.20\%$ for the combined sample.

Since a number of \mtau\ decay modes have not yet been precisely measured, particularly those modes which contain Cabibbo suppression 
factors, the branching fractions used as input to the \mc\ simulation come with an uncertainty which feeds into the total 
systematic error.
In order to evaluate this a weighted-sum of the \taubkgds\ is constructed using the \mc\ truth for \mtau\
\mc\ events passing the analysis selection criteria.

The branching fraction for the background mode $\tau^{-}\rightarrow K^{-} K^{+}\pi^{-}\nu_{\tau}$ is taken from a recent 
\babar\ collaboration measurement $\mathcal{B}(\tau^{-}\rightarrow K^{-} K^{+}\pi^{-}\nu_{\tau}) = (0.1346 \pm 0.0010 \pm 0.0036)\%$~\cite{Aubert:2007mh}.  
We predict the ratio 
$\mathcal{B}(\tau^{-}\rightarrow\pi^{-}K_{S}^{0} K^{0}_{L}\nu_{\tau})/\mathcal{B}(\tau^{-}\rightarrow K^{-} K^{+}\pi^{-}\nu_{\tau})$ 
to be $0.50\pm 0.05$ using an isospin relation~\cite{Finkemeier:1995sr}. 
This gives 
$\mathcal{B}(\tau^{-}\rightarrow\pi^{-}K_{S}^{0} K^{0}_{L}\nu_{\tau})=(0.0673 \pm 0.0070)\%$ 
which is consistent with the PDG 2006 value 
$\mathcal{B}(\tau^{-}\rightarrow\pi^{-}K_{S}^{0} K^{0}_{L}\nu_{\tau})=(0.112 \pm 0.030)\%$.
All other background mode uncertainties are taken from the PDG 2006~\cite{Yao:2006fs}.  The 
overall estimated uncertainty due to the \taubkgds\ is given by:
\begin{equation}
\label{eq:BRerror}
\Delta^{\scriptscriptstyle \mtau{\rm -bkg}} = \sqrt{\sum_{\scriptscriptstyle i} \left( w_{\scriptscriptstyle i} 
                                        \frac{\sigma_{\scriptscriptstyle i}} 
                                        {\BR_{\scriptscriptstyle i}} \right)^{\scriptscriptstyle 2}},
\end{equation}
where $w_{\scriptscriptstyle i}$ is the fraction of selected background $\tau$ events of mode $i$, 
$\BR_{\scriptscriptstyle i}$ is the branching fraction of mode $i$ and
$\sigma_{\scriptscriptstyle i}$ is the uncertainty of $\BR_{\scriptscriptstyle i}$.

Table~\ref{tab:TauBkgdsKpi0} shows the weights and uncertainties of the \taubkgd\ modes remaining in the selected 
$\tau^{-}\rightarrow K^{0}_{S}\pi^{-}\nu_{\tau}$ sample. 
The resulting uncertainty attributed 
to the \taubkgd\ modes ($\Delta^{\scriptscriptstyle \mtau{\rm -bkg}}$) is
estimated as $4.99\%$ on the number of background events and is consistent between each of the tagged samples.   
This leads to a $1.37\%$ systematic uncertainty on the branching fraction. 

\begin{table}[htbp]
\begin{center}
\caption{\footnotesize {\taubkgd\ uncertainties and their weights in the selected 
$\tau^{-}\rightarrow K^{0}_{S}\pi^{-}\nu_{\tau}$ sample.}}
\label{tab:TauBkgdsKpi0}

  \begin{tabular}{|c|c|c|}
    \hline
    Decay Channel   &   $w [\%] $
&$\frac{\sigma}{\mathcal{B}} [\%]$\\
\hline\hline
   $\tau^{-}\rightarrow\pi^{-} K_S^{0} \pi^{0}\nu_{\tau}$           & 24.24  &  10.53 \\
   $\tau^{-}\rightarrow K^{-}  K_S^{0} \pi^{0}\nu_{\tau}$           &  0.57  &  17.54  \\
   $\tau^{-}\rightarrow\pi^{-}\bar{K}^{0} K^{0}_{L}\nu_{\tau}$      & 41.16  &  10.38   \\
   $\tau^{-}\rightarrow\pi^{-}\bar{K}^{0} K^{0}_{S}\nu_{\tau} $     &  0.18  &   20.83  \\
   $\tau^{-}\rightarrow K^{-}\bar{K}^{0} \nu_{\tau}$                &  2.64  &   10.46  \\
   $\tau^{-}\rightarrow \pi^- \pi^0\nu_{\tau}$                      &  0.75  &   0.39   \\
   $\tau^{-}\rightarrow 2\pi^+\pi^-\nu_{\tau}$                      & 23.65  &   0.89   \\
   $\tau^{-}\rightarrow 2\pi^{-}\pi^{+}\pi^{0}\nu_{\tau}$           &  3.41  &   1.35   \\
   $\tau^{-}\rightarrow K^{-} K^{+}\pi^{-}\nu_{\tau}$               &  0.61  &   2.78   \\
   $\tau^{-}\rightarrow K^{-}\pi^{+}\pi^{-}\nu_{\tau}$              &  0.78  &  10.26  \\       
\hline
     \end{tabular}
\end{center}
\end{table}

        Table \ref{table:systSum} summarises the main sources of systematic uncertainty in the analysis.

        \begin{table}[htbp]     \begin{center}
        \caption{\footnotesize{Summary of the systematic uncertainties as they feed into the measurement of
$\mathcal{B}(\tau^{-}\rightarrow K^{0}\pi^{-}\nu_{\tau})$.}}
        \label{table:systSum}
        \begin{tabular}{|c|c|c|c|}
        \hline         
	Systematic                 	    	&       $e$-tag         &       $\mu$-tag               & Combined \\
        \hline\hline
        Tracking                        	&       0.58\%          &       0.58\%                  & 0.58\% \\
        $K_{S}^{0}$ Efficiency          	&       1.40\%          &       1.40\%                  & 1.40\% \\
        PID                             	&       1.45\%          &       1.68\%                  & 1.50\% \\
        $\mathcal{L}\times\sigma_{\tau\tau}$ 	&       0.83\%          &       0.83\%                  & 0.83\% \\
        Statistical efficiency error   		&       0.51\%          &       0.56\%                  & 0.38\% \\
        MC background statistics       		&       0.28\%          &       0.30\%                  & 0.20\% \\
        $\tau$ backgrounds              	&       1.37\%          &       1.37\%                  & 1.37\% \\
        Modelling efficiency            	&       0.37\%          &       0.37\%                  & 0.37\% \\
        \hline
        Total                           	&       2.73\%          &       2.87\%                  & 2.72\% \\
        \hline
        \end{tabular}
        \end{center}
        \end{table}

The relative systematic uncertainty on the branching fraction includes the individual sources described 
above when propagated through equation~(\ref{eq:BR}) taking into account correlations between the different components 
considered.  The total systematic uncertainty on the branching fraction measurement is $2.73\%$, $2.87\%$ and $2.72\%$ 
for the \etag, \mutag\ and combined samples, respectively.

%% file: sections/branchingRatio.tex
\begin{figure}[htbp]
\begin{center}
\subfigure{\label{fig:LogScale}\includegraphics[width=27pc,height=22.5pc]{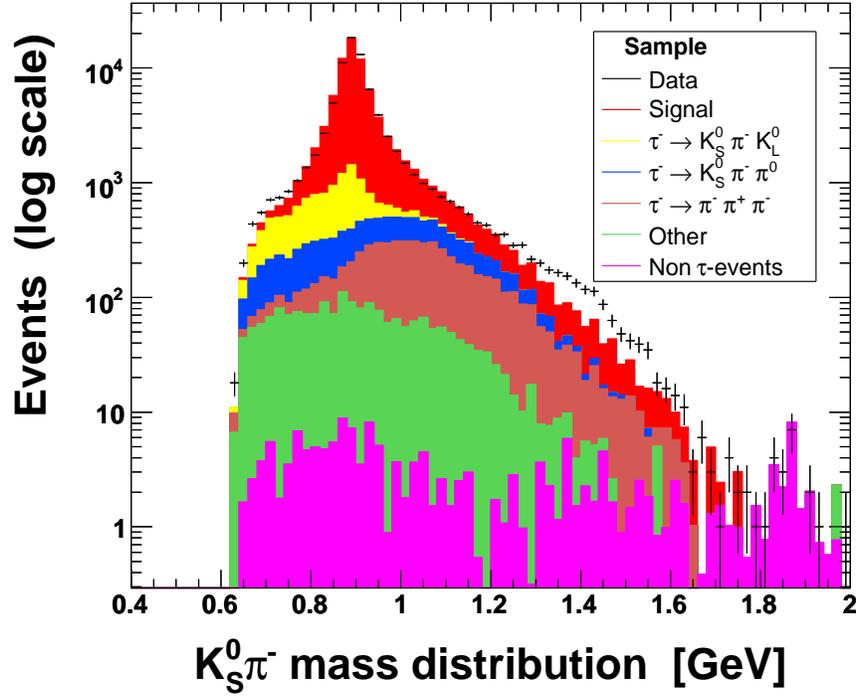}}
\subfigure{\label{fig:Scale}\includegraphics[width=27pc,height=22.5pc]{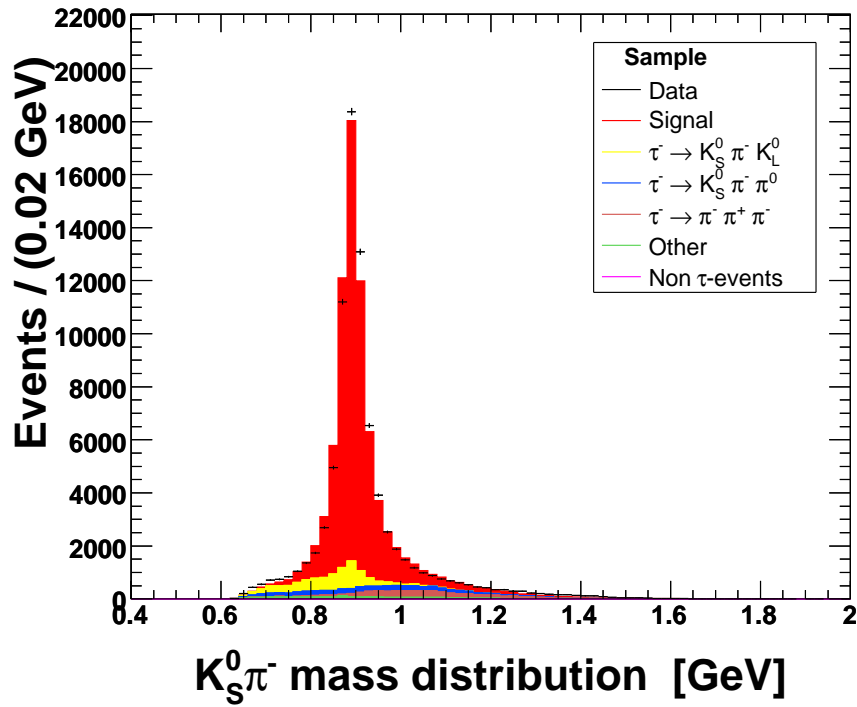}}
\caption{\footnotesize{Reconstructed $\tau^{-}\rightarrow K^{0}_{S}\pi^{-}\nu_{\tau}$ mass distribution for 
the combined (\etag+\mutag) sample using (left) logarithmic scale and (right) linear scale.  The MC signal 
branching fraction 
is set to the measured value from this analysis.}}
\label{fig_masscomparison}
\end{center}
\end{figure}

Figure\,\ref{fig_masscomparison} shows the invariant mass spectrum of the selected data and MC $K^{0}_{S}\pi^{-}$ 
candidates
in the combined (\etag+\mutag) sample after all the analysis requirements, including efficiency corrections.
In this plot, the signal MC sample is scaled using the branching fraction measured in this analysis: 
$\mathcal{B}(\tau^{-}\rightarrow K^{0}_{S}\pi^{-}\nu_{\tau}) = 0.420\%$.  The data and MC sample disagreements
can be ascribed to inadequate modelling of the $K^{*}$ resonances in the MC simulation.  There is some evidence 
that the $K^{*}(892)$ mass has been underestimated in the MC simulation and of the existence of $K^{*}_{0}(800)$ 
and $K^{*}_{0}(1430)$ 
resonances which are not modelled~\cite{BellePaper}.  However these discrepancies do not affect our result for 
the branching ratio.

The final result for $\mathcal{B}(\tau^{-}\rightarrow \bar{K^{0}}\pi^{-}\nu_{\tau})$ is estimated using the combined 
$e$-tag + $\mu$-tag sample, where
the total number of events observed, estimated background level and efficiency are derived from the two tagged
samples. 
As a cross-check, $\mathcal{B}(\tau^{-}\rightarrow \bar{K^{0}}\pi^{-}\nu_{\tau})$ is calculated for each tagged 
sample separately and the results obtained are in excellent agreement with the combined result.

The branching fraction $\mathcal{B}(\tau^{-}\rightarrow \bar{K^{0}}\pi^{-}\nu_{\tau})$ is estimated by

\begin{equation}
\mathcal{B}(\tau^{-}\rightarrow \bar{K^{0}}\pi^{-}\nu_{\tau}) = \frac{1}{2\Ntautau} \frac{N_{\scriptscriptstyle{\rm data}}
               - N_{\scriptscriptstyle{\rm bkg}}}{\SigEff},
\label{eq:BR}
\end{equation}

\noindent
where $\Ntautau$ is the total number of \tautau\ pairs in the real data, $N_{\scriptscriptstyle{\rm data}}$ is the number
of selected events in real data, $N_{\scriptscriptstyle{\rm bkg}}$ is the number of background events estimated from \mc\ sample and $\SigEff$ is the signal efficiency as defined in equation~(\ref{eq:SigEff}).
The total number of \tautau\ pairs in the data is given by

\begin{equation}
\Ntautau = \sigma_{\scriptscriptstyle \mtau} \lum_{\scriptscriptstyle {\rm data}} = (353.4 \pm 2.3)\times 10^{6}, 
\end{equation}

\noindent
where $\sigma_{\scriptscriptstyle \mtau}$ is the \tautau\ production \crosssection\ at \babar\ (\ie\ $0.919 \pm 0.003 \nb$) and 
$\lum_{\scriptscriptstyle {\rm data}}$ is the (integrated) real data luminosity (\ie\ $384.6 \pm 2.2 \invfb$).

Table~\ref{tab:kpi0BRcontributions} gives the numbers of real and simulated data events passing the selection criteria
that feed into the $\mathcal{B}(\tau^{-}\rightarrow \bar{K^{0}}\pi^{-}\nu_{\tau})$ calculation.   
A result for the measurement of the branching fraction $\mathcal{B}(\tau^{-}\rightarrow
\bar{K^{0}}\pi^{-}\nu_{\tau})$ is given in Table~\ref{tab:kpi0BR}.

\begin{table}[htbp]
\begin{center}
\caption{\footnotesize{Numbers of data and \mc\ events remaining in the selected samples corresponding to the 
$384.6 \invfb$ (integrated) real data luminosity.  The signal \mc\ has been scaled with the PDG 2006 value~\cite{Yao:2006fs}.}} 
\label{tab:kpi0BRcontributions}
\begin{tabular}{@{}llll}
\hline
       Data          &       \etag           &         \mutag        &           Combined \\
\hline
Real                 & 47092    $\pm$ 217        &  36641   $\pm$ 191         &  83733   $\pm$ 289  \\
Signal MC 	     & 39445    $\pm$ 193        &  30749   $\pm$ 176         &  70194   $\pm$ 261  \\
\tautau\ background  & 9942     $\pm$ 92         &  7645    $\pm$ 88          &  17587   $\pm$ 131  \\  
\uds                 & 8.9      $\pm$ 3.4        &  65.2    $\pm$ 7.9         &  74.1    $\pm$ 8.5  \\
\ccbar               & 45.5     $\pm$ 5.0        &  43.5    $\pm$ 4.8         &  89.1    $\pm$ 6.9  \\
\BB                  & 3.4      $\pm$ 1.1        &  3.7     $\pm$ 1.2         &  7.1     $\pm$ 1.6  \\
$\mu^{+}\mu^{-}$     & 0        $\pm$ 0          &  14.2    $\pm$ 3.7         &  14.2    $\pm$ 3.7  \\
\hline
\end{tabular}
\end{center}
\end{table}

\begin{table}[htbp]
\begin{center}
\caption{\footnotesize{\babar\ preliminary $\mathcal{B}(\tau^{-}\rightarrow \bar{K^{0}}\pi^{-}\nu_{\tau})$ measurements.}}
\label{tab:kpi0BR}
\begin{tabular}{@{}ll}
\hline
   Sample     &    $\mathcal{B}(\tau^{-}\rightarrow \bar{K^{0}}\pi^{-}\nu_{\tau})$ [\%] \\
\hline
\etag         & $ 0.840 \pm 0.005 \, \stat \pm 0.023 \, \syst$ \\
\mutag        & $ 0.840 \pm 0.006 \, \stat \pm 0.024 \, \syst$ \\
Combined      & $ 0.840 \pm 0.004 \, \stat \pm 0.023 \, \syst$ \\
\hline
\end{tabular}
\end{center}
\end{table}

\begin{figure}[htbp]
\begin{center}
\includegraphics[width=27pc,height=27pc]{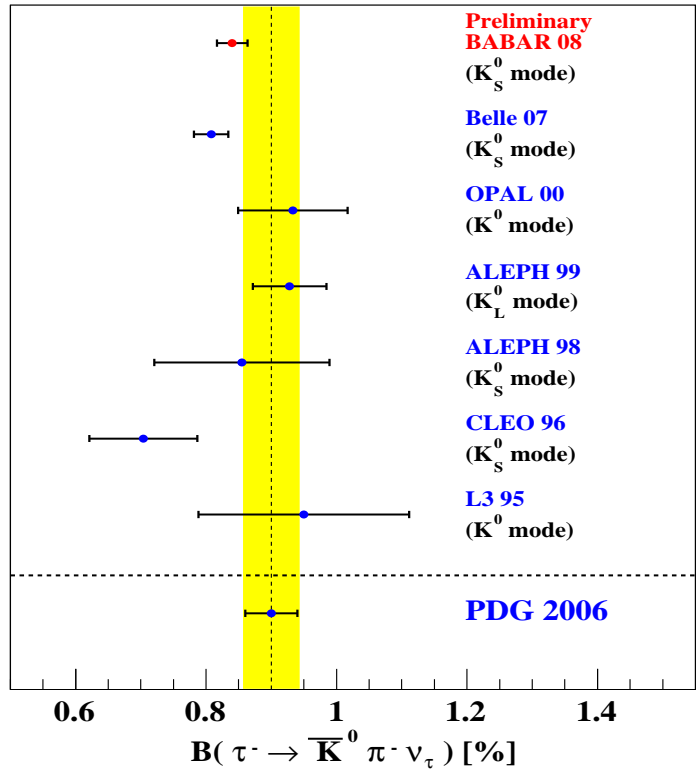}
\caption{\footnotesize{Previously published measurements of $\mathcal{B}(\tau^{-}\rightarrow \bar{K^{0}}\pi^{-}\nu_{\tau})$.
The Belle result and the result from this analysis do not contribute to the PDG 2006 average.
The \babar\ preliminary result provides the world's most precise measurement to date:
$\mathcal{B}(\tau^{-}\rightarrow \bar{K^{0}}\pi^{-}\nu_{\tau}) = \left( 0.840 \pm 0.004 \, \stat \pm 0.023 \, \syst \right)\%$.}}
\label{fig_brcomparison}
\end{center}
\end{figure}

Figure~\ref{fig_brcomparison} shows previously published measurements of $\mathcal{B}(\tau^{-}\rightarrow
\bar{K^{0}}\pi^{-}\nu_{\tau})$.  It can be seen that this \babar\ preliminary result is the world's most precise measurement.

%% file: sections/summary.tex
Using $384.6 \invfb$ of \epem\ collision data produced by the \pep2\ collider and recorded by the \babar\ detector, 
we obtain the preliminary result: 
\begin{eqnarray}
\nonumber
\lefteqn{\mathcal{B}(\tau^{-}\rightarrow \bar{K^{0}}\pi^{-}\nu_{\tau}) =} 
\\ && \mbox{}
\left( 0.840 \pm 0.004 \, \stat \pm 0.023 \, \syst \right) \%.
\end{eqnarray}
This result is the most precise measurement of this branching fraction to date and is consistent with the world average.

%% file: acknowledgements.tex
We are grateful for the 
extraordinary contributions of our \pep2\ colleagues in
achieving the excellent luminosity and machine conditions
that have made this work possible.
The success of this project also relies critically on the 
expertise and dedication of the computing organizations that 
support \babar.
The collaborating institutions wish to thank 
SLAC for its support and the kind hospitality extended to them. 
This work is supported by the
US Department of Energy
and National Science Foundation, the
Natural Sciences and Engineering Research Council (Canada),
the Commissariat \`a l'Energie Atomique and
Institut National de Physique Nucl\'eaire et de Physique des Particules
(France), the
Bundesministerium f\"ur Bildung und Forschung and
Deutsche Forschungsgemeinschaft
(Germany), the
Istituto Nazionale di Fisica Nucleare (Italy),
the Foundation for Fundamental Research on Matter (The Netherlands),
the Research Council of Norway, the
Ministry of Education and Science of the Russian Federation, 
Ministerio de Educaci\'on y Ciencia (Spain), and the
Science and Technology Facilities Council (United Kingdom).
Individuals have received support from 
the Marie-Curie IEF program (European Union) and
the A. P. Sloan Foundation.